# MPGDs in Compton imaging with liquid-xenon

**Samuel Duval**[a*], **Amos Breskin**[b], **Herve Carduner**[a], **Jean-Pierre Cussonneau**[a], **Jacob Lamblin**[a], **Patrick Le Ray**[a], **Eric Morteau**[a], **Tugdual Oger**[a], **Jean-Sebastien Stutzmann**[a] **and Dominique Thers**[a].

[a] *Subatech, Ecole des Mines, CNRS/IN2P3 and Université de Nantes,*
*44307 Nantes, France*

[b] *Departement of Physics, The Weizmann Institute of Science,*
*76100 Rehovot, Israel*
*E-mail*: `samuel.duval@subatech.in2p3.fr`

ABSTRACT: The interaction of radiation with liquid xenon, inducing both scintillation and ionization signals, is of particular interest for Compton-sequences reconstruction. We report on the development and recent results of a liquid-xenon time-projection chamber, dedicated to a novel nuclear imaging technique named "3γ imaging". In a first prototype, the scintillation is detected by a vacuum photomultiplier tube and the charges are collected with a MICROMEGAS structure; both are fully immersed in liquid xenon. In view of the final large-area detector, and with the aim of minimizing dead-zones, we are investigating a gaseous photomultiplier for recording the UV scintillation photons. The prototype concept is presented as well as preliminary results in liquid xenon. We also present soft x-rays test results of a gaseous photomultiplier prototype made of a double Thick Gaseous Electron Multiplier (THGEM) at normal temperature and pressure conditions.

---

[*] Corresponding author

# Contents

## 1. Introduction

Positron emission tomography is a nuclear medical imaging technique mainly used in oncology which allows the follow-up of a radiotracer biodistribution into a targeted tissue. This technique permits the visual interpretation of tissues or organs metabolism. However, some factors lead to quantification uncertainties like patient motion, γ-rays scattering, attenuation of radiation by the body, tomography reconstruction, etc [1]. Among these factors we propose a method for reducing artifacts due to image reconstruction. For this purpose each point of disintegration has to be localized individually in three dimensions, in real time. Few years ago, a novel imaging technique was initiated at Subatech [2]. In a standard PET device the $\beta^+$ emitter localization is based on the detection in coincidence of the two back-to-back annihilation γ-rays. Hence, we only know that the disintegration occurs in the patient along the line-of-response (LOR). Using a specific radioisotope emitting a γ just after the $\beta^+$ decay permits an event-by-event detection of the three photons. The direction of the additional emitted γ-ray is inferred from the Compton kinematics as shown in figure 1. The first hit energy leads to the determination of the aperture angle θ of the Compton cone and the position of the emitter can then be measured by calculating the intersection between the cone and the LOR, measured by a classical PET device [3].

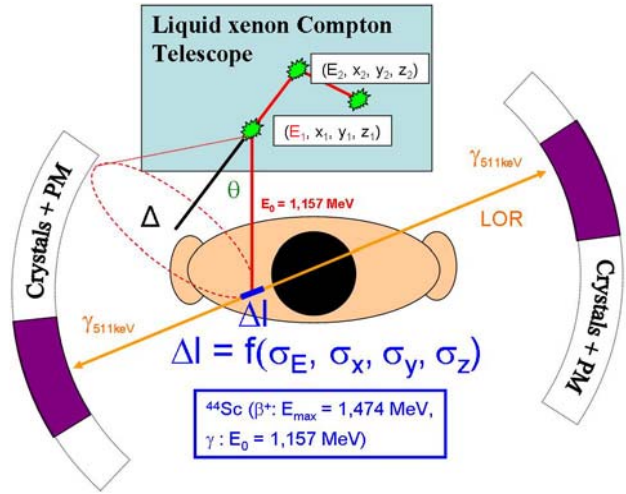

Figure 1: Principle of the 3γ imaging with a Compton telescope coupled to a PET device.

## 2. XEMIS (Xenon Medical Imaging System)

### 2.1 Experimental Setup

A liquid-xenon time-projection chamber (LXe TPC) has been built in order to validate the 3 γ imaging concept. The experimental setup is shown in figure 2 (left). It consists of a TPC inserted within a cryostat maintained at around -110ºC under a pressure of 1.3 bar, a purification circuit equipped with a getter (MonoTorr® Phase II PURIFIER PS4-MT3/15-R/N-1/2) and an electronic bench. The xenon is required to be as pure as possible to prevent charge carriers capture by electronegative impurities ($H_2O$, $O_2$) released by construction materials into the liquid.

The TPC (length: 12 cm, diameter: 3.6 cm) was composed of a PMT (Hamamatsu R5900-06AL12S-ASSY), for chamber triggering, and of a MICROMEGAS [4] device for charge carriers readout - both fully immersed in liquid xenon. The UV-light emitted by γ photons interacting within the TPC volume was guided toward the PMT by a PTFE wall; the simultaneously induced charge carriers were drifted in the liquid toward the MICROMEGAS through an electric field of 2kV/cm. The micromesh of the MICROMEGAS was used as a Frisch grid; it was placed 50 μm above a square anode of 6.5 $cm^2$. The micromesh characteristics were: hole spacing ("s"): 60 μm, hole diameter ("d"): 30 μm, copper thickness ("t"): 5 μm and pillars height ("p"): 50μm (figure 2 (right)). A collimated $^{22}$Na ($E_{max\ \beta+}$ = 545 keV, $E_\gamma$: 1.257 MeV) source was placed in front of the TPC. We observed 511 keV γ-rays, triggering the chamber in coincidence with a CsI crystal coupled to a PMT at the opposed side. Measurements of electron life-length were done by recording the energy spectrum corresponding to the 511 keV γ-rays interacting at different depths of the TPC's sensitive volume. An example of 511keV recorded event is shown in figure 3. The electron life-length was deduced from the fit of the exponential decay of the pulse-height with depth, as function of purification time. The grid transparency to charges was measured at variable drift fields, with a fixed induction field, recording shifts of the photoelectric peak.

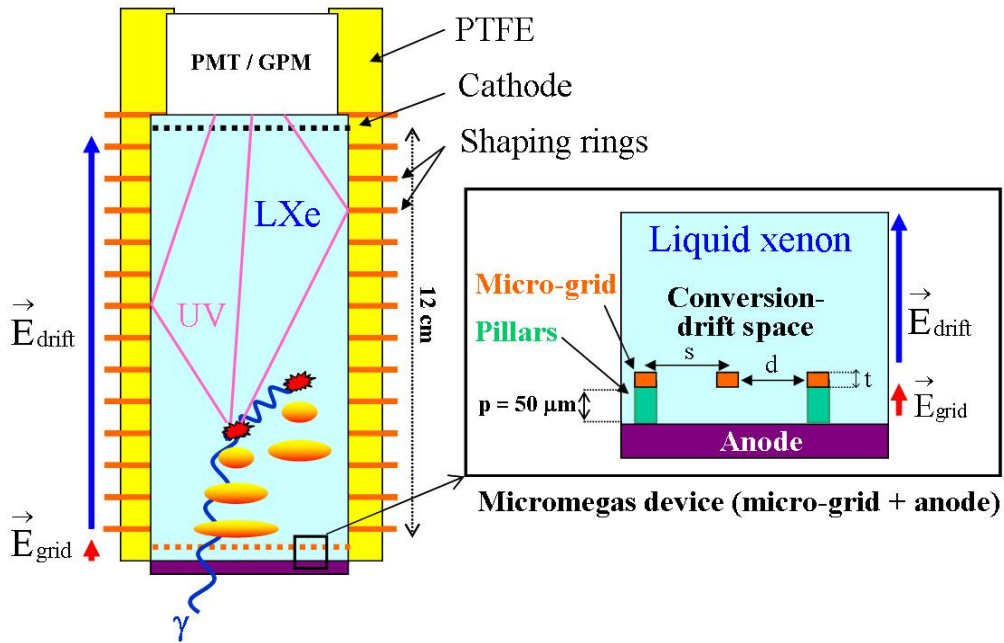

Figure 2: Schematic drawing of the experimental setup: the TPC (left) and zoom shot of the charge readout device (right).

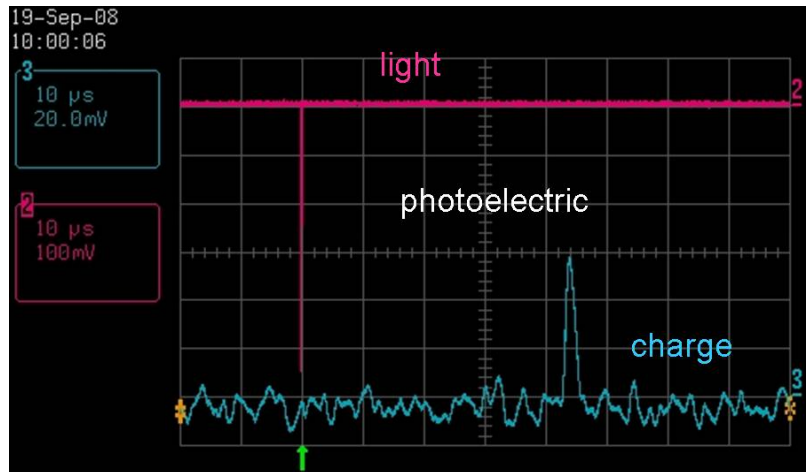

Figure 3: Oscilloscope snapshot of a photoelectric interaction: light pulse (top), charge pulse (bottom).

### 2.2 Results

Figure 4 shows the evolution of electron life-length as a function of purification time. This underlines the high purity of the liquid xenon. Indeed, after one month of circulation, charge carriers could drift along one meter without being captured by electronegative impurities. Another important and noticeable result (Figure 5) is the trend of the micromesh transparency in LXe which is quite similar to that of standard gases. It reaches a plateau for a field ratio of 50.

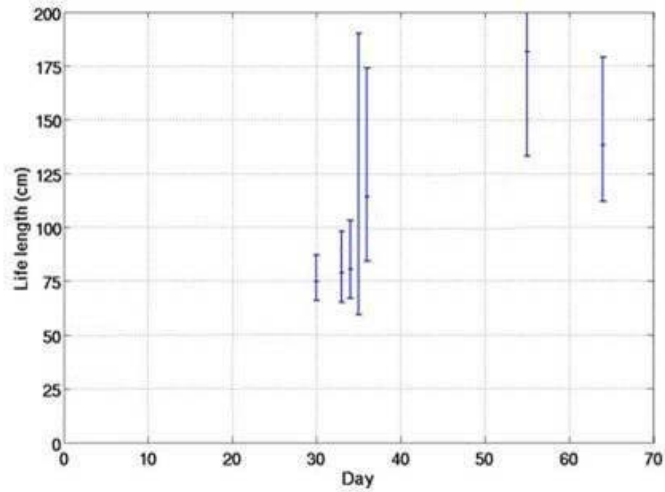

Figure 4: Electrons life-length as a function of purification time of liquid xenon.

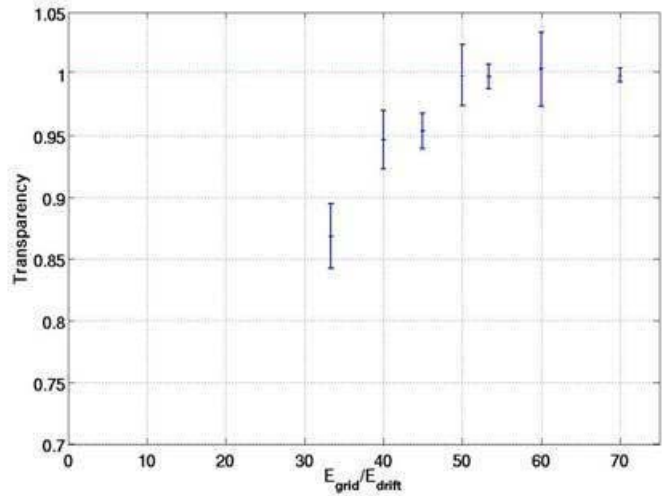

Figure 5: Micromesh (s: 60 μm, d: 30 μm, t: 5 μm and p: 50 μm) charge-transparency measurements as function of $E_{grid}/E_{drift}$ ratio in liquid xenon (figure 2).

### 3. A large cryogenic UV-GPM for XEMIS 2

In view of a future large-size LXe TPC development for small-animal medical imaging we have been pursuing the development of a large UV-gaseous photomultiplier (GPM). Such device that can be economically produced over large area combines a CsI UV-sensitive photocathode and a gas-avalanche electron multiplier [5]. It is expected to provide an efficient large homogenous sensitive area, with no dead-layers, with position sensitivity ("local triggering") and efficiency to low light levels – down to single photons. First experimental tests of GPM prototypes are presented in the following subsections.

#### 3.1 A local triggering

In case of a large noble gas TPC volume, the whole ionization readout plan (x,y) is busy during the charge carriers drift along z axis. We propose to reduce the detector occupancy applying a

"local trigger" concept consisting of opening a cylindrical sensitive volume of which the revolution axis passes through the center of gravity of the LXe scintillation signal. That requires that the scintillation signal must be the least possible interaction position-dependent. As illustrated in figure 6a), the higher the radius of the cylinder, the better is the probability of including the two first hits of the Compton sequence is. "% of triggered events" on the vertical axis of figure 6a) are events in which the two first hits (interactions) are inscribed within a sensitive volume of a given radius. Simulations of scintillation signal collection were carried out with GEANT4 for γ-ray interactions and a Monte-Carlo code for optical-photons transport with ROOT framework [6]. These simulations underlined that a large GPM with a $MgF_2$ window, a reflective CsI photocathode (e.g. QE~30% at 170nm [7]) and a segmented anode (81 pads of 2.8×2.8 $cm^2$) is less interaction-location dependent compared to a readout plane made of 81 (2.8×2.8 $cm^2$) PMTs. The number of triggered events increases sharply as a function of the cylinder radius. Hence, only the corresponding section which faces the cylinder is triggered. In figure 6b) we can see that both the large GPM and the PMTs array similarly localize the two first hits' center of gravity. Nevertheless, the GPM is an economical, simple and elegant solution.

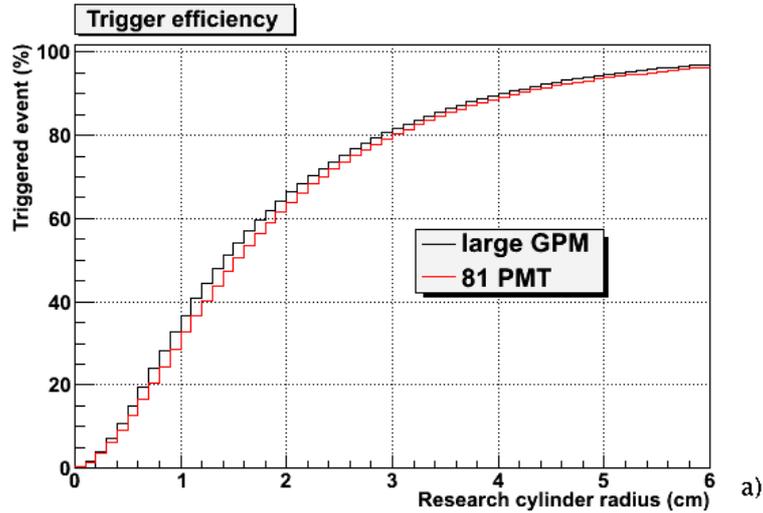

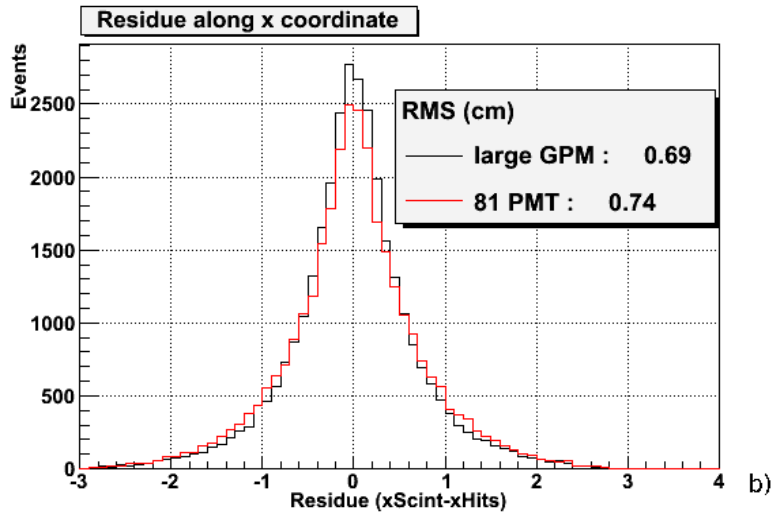

Figure 6: a) Trigger efficiency for 1.157 MeV multi-hit γ-rays for a single large GPM with a segmented anode or an array of 81 PMTs as a function of the radius cylinder. b) Residue along x axis, between the scintillation center of gravity (xScint) and the two first hits center of gravity (xHits).

### 3.2 Materials and Methods

A cylindrical GPM prototype with a 32 mm diameter anode was especially designed for cryogenic applications and pressure stresses. Two different setups were tested. The first setup was composed of two FR4 made thick GEMs (THGEM) [8] as shown in figure 7. The THGEM geometry was: 400 μm thickness ("t"), with 300 μm hole diameter ("d"), a rim size at the hole edge of 50 μm ("r") and a hole spacing of 700 μm ("s"). The conversion, transfer and induction gaps were 4 mm. The second setup was made of a single THGEM (same characteristics as above) and a copper made MICROMEGAS [4] of dimensions: s: 60 μm, d: 30 μm, t: 5 μm and p: 50 μm. The conversion gap was 8 mm and the transfer gap was 4 mm (figure 8). For both setups gain measurements were carried out with $^{55}$Fe soft X-rays in a Ne/5%CH$_4$ mixture at room temperature.

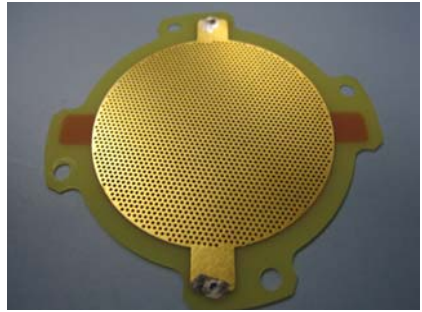

Figure 7: Photograph of a THGEM

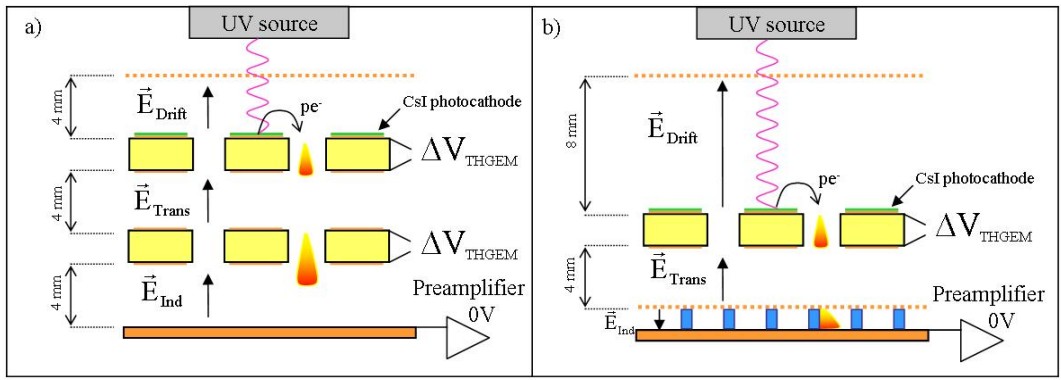

Figure 8: Schematic drawings of the Gaseous Photomultiplier set-up. a) Double-THGEM and b) THGEM followed by a MICROMEGAS. Both were tested with soft x-rays without CsI photocathode.

### 3.3 Results

As shown in figures 9 and 10 both setups yielded high gains ~5x10$^4$ with the double THGEM and ~2x10$^6$ with the THGEM/MICROMEGAS. It should be noted that recent extensive studies of CsI-coated THGEM photon detectors [9] demonstrated high efficiency of photoelectron detection and yielded somewhat higher gains ($3\times10^5$) at relatively low operation voltages.

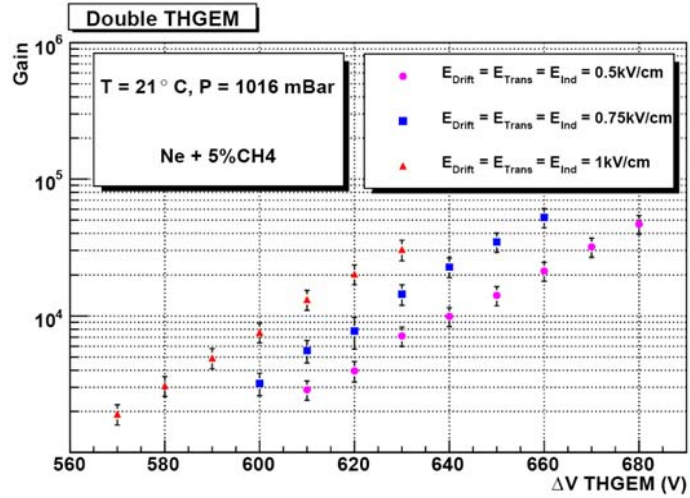

Figure 9: Gain curves obtained with soft x-rays in the double-THGEM (t: 400 μm, d: 300 μm, r: 50 μm, s: 700 μm) detector operated in Ne/5%CH4 mixture at room temperature.

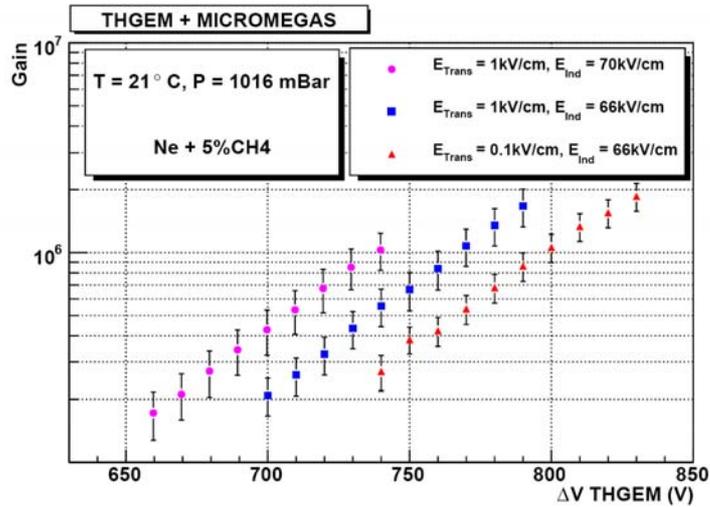

Figure 10: Gain curves obtained with a THGEM (t: 400 μm, d: 300 μm, r: 50 μm, s: 700 μm) followed by a MICROMEGAS (s: 60 μm, d: 30 μm, t: 5 μm, p: 50 μm) detector operated at room temperature in Ne/5%CH$_4$.

## 4. Conclusions

Previous simulation studies [2][10] demonstrated that 3 γ imaging is a promising imaging technique and unique in terms of real-time imaging. First experimental results (electron life-length measurements) showed the excellent purification capability of our facility over around two months. Moreover, for the first time we demonstrated the functioning of a MICROMEGAS device in a Frisch grid mode immersed in liquid-xenon.

While first results were measured with a vacuum PMT, intensive R&D is directed towards the development of a large-area gaseous photomultiplier, capable of operation in cryogenic

–

conditions. Experimental results and first characterizations in Ne/5%CH$_4$ are encouraging although the setups were not optimized. This mixture permits to attain high gains with low voltages applied to the THGEM, which is important for its stability. Furthermore neon and methane remain gaseous at 165 K at 1.3 bar pressure, which will be the experimental conditions inside the Compton telescope. Moreover cesium iodide reflective photocathodes have shown good photoelectron extraction and transmission into THGEM holes in Ne/CH$_4$ mixtures due to a relatively low electron backscattering [11]. This type of photocathode was demonstrated to function in cryogenic conditions [12] and so did the THGEM [13]. Finally, the THGEM-MICROMEGAS device should have a low ion backflow reduction [14], of importance for the operation stability and photocathode lifetime. Investigations of both GPM configurations at cryogenic conditions will include long-term gain stability studies at different operation conditions.

## Acknowledgements

This work was partly supported by the Israel Science Foundation, grant No 402/05, by a France-Israel cooperation grant of the Israel Ministry of Science and the French Ministry of Education, as well as by the region of Pays de la Loire, France. A. Breskin is the W.P. Reuther Professor of Research in The Peaceful Use of Atomic Energy.

## References

[1] I. Buvat, *Quantification in emission tomography: Challenges, solutions and performances*, 2007 NIMA 571, 10-13

[2] C.Grignon et al., *Nuclear medical imaging using $\beta^+\gamma$ coincidence from Sc$^{44}$ radio-nuclide with liquid xenon as detection medium,* 2007 NIMA 571, 142-145

[3] Centre National de la Recherche Scientifique et (CNRS) et Ecole des Mines de Nantes, *Telescope Compton au xenon liquide*, PATENT 2 926 893

[4] Y. Giomataris et al., *MICROMEGAS: a high-granularity position-sensitive gaseous detector for high particle-flux environments*, 1996 NIMA, 376, 29-35

[5] R. Chechik and A. Breskin; *Advances in Gaseous Photomultipliers*; Nucl. Instr. Meth. A595 (2008) 116-127

[6] ROOT, an object-oriented data analysis framework. ROOT website

[7] A. Breskin, NIM A A371 (1996) 116

[8] A. Breskin, R. Alon, M. Cortesi, R. Chechik, J. Miyamoto, V. Dangendorf, J. Maja, J.M.F. Dos Santos; *A concise review on THGEM detectors*; Nucl. Instr. Meth. A598(2009)107

[9] M. Cortesi et al. 2009 JINST, in press; arXiv:0905.2916

[10] C. Grignon, *Etude et développement d'un télescope Compton au xénon liquide dédié a l'imagerie médicale fonctionnelle*, PhD Thesis of Nantes University, 2007, N° ED 366-348

[11] Cortesi Ne paper, J. Escada, these proceedings

[12] L. Periale et al., *Development of gaseous detectors with solid photocathodes for low-temperature applications*, 2004 NIMA 535, 517-522

–

[13] A. Bondar et al. 2008_JINST_3_P07001

[14] P. Colas et al., *Ionbackflow in the Micromegas TPC for the future linear collider*, 2004, NIMA 535, 226-230